\documentclass[11pt]{article}
  
\usepackage{amsmath,amssymb}

\usepackage{epsfig}  
\usepackage{graphicx}               
\usepackage{url}
\usepackage{hyperref}

\newcommand{\nc}{\newcommand}  

\nc{\beq}{\begin{equation}}  
\nc{\eeq}{\end{equation}}  
\nc{\beqa}{\begin{eqnarray}}  
\nc{\eeqa}{\end{eqnarray}}  
\nc{\bea}{\begin{eqnarray}}  
\nc{\eea}{\end{eqnarray}}  
\nc{\ra}{\rightarrow}  
\nc{\lsim}{\begin{array}{c}\,\sim\vspace{-21pt}\\< \end{array}}  
\nc{\gsim}{\begin{array}{c}\sim\vspace{-21pt}\\> \end{array}}  
\nc{\Tr}{{\rm Tr}}
\nc{\slsh}{\slash\hspace*{-0.22cm}}

\def\be{\begin{equation}}
\def\ee{\end{equation}}
\def\bea{\begin{eqnarray}}
\def\eea{\end{eqnarray}}
\def\bit{\begin{itemize}}
\def\eit{\end{itemize}}
\nc{\er}[1]{(\ref{eq:#1})}

\nc{\ttbar}{t\bar t}

\newcommand{\gev}{{\rm GeV}}

\def\ltap{\ \raise.3ex\hbox{$<$\kern-.75em\lower1ex\hbox{$\sim$}}\ }
\def\gtap{\ \raise.3ex\hbox{$>$\kern-.75em\lower1ex\hbox{$\sim$}}\ }

\title{  
\vspace*{-2.3cm}  
\begin{flushright}  
\normalsize{  
SLAC-PUB-14509 
  }  
\end{flushright}  
\vspace{1.5cm}  
\Large  
\textbf{
Composite Octet Searches with Jet Substructure 
}\vspace*{1.0cm}   
}

\author{Yang Bai$^a$
and Jessie Shelton$^{b}$ 
\vspace{5mm}
\\
${}^{a}$\normalsize\emph{SLAC National Accelerator Laboratory, 2575 Sand Hill Road, Menlo Park, CA 94025, USA} 
    \vspace{1mm} \\
${}^{b}$\normalsize\emph{Department of Physics, Sloane Laboratory, Yale University, New Haven, CT, 06520, USA}
}

\date{}

\begin{document}  
\setcounter{page}{0}  
\maketitle  

\vspace*{1cm}  
\begin{abstract} 
Many new physics models with strongly interacting sectors predict a mass hierarchy between 
the lightest vector meson and the lightest pseudoscalar mesons. We examine the power of jet 
substructure tools to extend the 7 TeV LHC sensitivity to these new states for the case of 
QCD octet mesons, considering both two gluon and two b-jet decay modes for the pseudoscalar 
mesons. We develop both a simple dijet search using only the jet mass and a more sophisticated 
jet substructure analysis, both of which can discover the composite octets in a dijet-like 
signature.  The reach depends on the mass hierarchy between the vector and pseudoscalar mesons.  
We find that for the pseudoscalar-to-vector meson mass ratio below approximately 0.2 the simple 
jet mass analysis provides the best discovery limit; for a ratio between 0.2 and the QCD-like value 
of 0.3, the sophisticated jet substructure analysis has the best discovery potential; for a ratio 
above approximately 0.3, the standard four-jet analysis is more suitable. 
\end{abstract}  
\thispagestyle{empty}  
\newpage  
  
\setcounter{page}{1}

\baselineskip18pt   

\section{Introduction}
\label{sec:intro}

With the advent of the LHC, the electroweak scale is being probed at
last.  However, the startlingly good agreement of the standard model
(SM) with precision flavor and electroweak measurements begs the
question: does TeV-scale physics substantially modify the SM story of
electroweak symmetry breaking (EWSB)?  Hints from low-energy data
suggest that physics Beyond the Standard Model (BSM), if it has to do
with flavor, EWSB, or leptons, may be heavy, while new physics at the
TeV scale may more comfortably be hadrophilic, and not obviously
related to the mysteries of EWSB.  Such a situation poses theoretical 
puzzles.  However, the possibility is one which should be seriously considered:
Nature has a track record of handing us 
particles for which we have no obvious need.  There are many open
possibilities for interesting BSM physics at the TeV scale, many of
which present large signals for the early LHC.

One such possibility is a new non-Abelian gauge interaction $G$ which
confines at scales $\Lambda_G\gtap$ TeV. To access this sector, some new particles must
couple to both the $G$ sector and the SM; the simplest
possibility is a fermion species $\Psi$ which transforms under both
$G$ and the SM gauge groups.  If $\Psi$ is in a vector representation 
($\Psi_L$ and $\Psi_R$ have the same quantum numbers) of the SM
gauge group and does not have large mixings with SM fermions, SM precision observables are unaffected.  For 
$\Psi$ charged under QCD, after the group $G$ confines, the spectrum of
$G$-hadrons will include several new colored states~\cite{Kilic:2008pm,  Kilic:2009mi, Bai:2010qg}.

This simple scenario naturally gives rise to signals which are
particularly well-suited to study at the LHC: new colored states with
masses at  the TeV scale or below.  In particular, the $G$ sector will
typically contain a relatively light color-octet vector meson $\rho^a_G$ 
with $m_{\rho_G}\sim \Lambda_G \gtap $ TeV, and colored pseudo-Nambu-Goldstone 
bosons (pNGB's) $\pi^a_G$ with masses $m_{\pi_G}$ parametrically lighter 
than $\Lambda_G$.  These states are particularly important for the LHC 
phenomenology of a new confining interaction, as the $\rho^a_G$ can be 
resonantly produced through its mixing with the gluon, while the $G$-pions $\pi^a_G$, 
as the lightest colored particles in the $G$ sector, have the largest production
cross sections.\footnote{Here we take the bare masses of the new fermion species
to be negligible in comparison with the confinement scale, $m_\Psi \ll
\Lambda_G$.  One can introduce a Peccei-Quinn symmetry to forbid the bare 
fermion masses~\cite{Bai:2010qg}.  The opposite situation, $\Lambda_G \ll m_\Psi$, leads to
quirks, with very different phenomenologies~\cite{Kang:2008ea}.}

While the production cross sections for these light composite states
are large, the dominant decays of the $\pi^a_G$ and the $\rho^a_G $  yield
all-hadronic final states, and large QCD backgrounds can make
discovery challenging \cite{Kilic:2008ub}.  This is especially true of
the vector resonance, $\rho^a_G$, which naturally has large
branching fractions to other new states, rather than back to dijets,
thus leading to multijet final states which can be challenging to
separate from QCD backgrounds.  Recent advances in jet substructure
have extended LHC sensitivity to both SM \cite{Butterworth:2002tt, Butterworth:2008iy, jss-sm} and BSM \cite{jetmass, jss-bsm} 
signals in otherwise challenging multi-jet final states.  Here, we will demonstrate the power of simple
jet substructure techniques to improve LHC discovery sensitivity to
colored resonances.  The hierarchy of mass scales
$m_{\pi_G}/m_{\rho_G}$ means that the $G$-pions produced in the decay
of a $\rho_G^a$ are boosted.  The subsequent decay $\pi_G\to gg,\,b\bar
b$ can be extracted from the large QCD background using a simple and
flexible $G$-pion tagger, which distinguishes the perturbative
$G$-pion decay from the shower structure of a QCD jet.
Our main motivation is a confining gauge group, as this scenario
naturally generates the hierarchy of mass scales which necessitates a
substructure analysis, but the techniques presented here are useful
for any theory with a colored vector resonance (``coloron'' \cite{Hill:2002ap, Dobrescu:2007yp, Bai:2010dj}) or colored axial vector
resonance (``axigluon'' \cite{Pati:1974zv, Hall:1985wz, Frampton:1987dn}) with new colored daughters.

The organization of this paper is as follows.  In Section~\ref{sec:model} we
introduce a simplified model capturing the dynamics of interest and
discuss the parameter space of the theory.  In Section~\ref{sec:discovery} we introduce
two simple jet substructure searches and show the discovery reach of
the 7 TeV LHC for both.  Section~\ref{sec:conclusions} contains our conclusions, 
and in Appendix~\ref{sec:vectorconfinement} we provide a more extended discussion of how our
simplified model fits into the Lagrangian of a generic confining
sector.

\section{A simple model for spin-1 and spin-0 composite octets}
\label{sec:model}

One generic possibility for physics above the electroweak scale is a
new gauge interaction $G $ which confines at a scale $\Lambda_G$ above the electroweak scale.  New fermion species $\Psi$ which transform under $G$
then are not observed in isolation at colliders, but rather in bound
states which are singlets of $G$, which we call $G$-hadrons.  This
idea is hardly new: technicolor is one example of such a model.
Unlike technicolor, however, we do not necessarily imagine here that
the chiral condensation of $G $-fermions is responsible for
electroweak symmetry breaking.

In general the new confining gauge group will result in a rich
spectrum of $G$-hadrons with a range of SM quantum numbers.  In the
spirit of Simplified Models~\cite{Alves:2011wf}, we introduce a
simplified model which succinctly captures the most relevant dynamics
for discovery at hadron colliders, especially in dijet or multi-jet
final states.  A discussion of how this simplified model maps onto
such well-motivated extensions of the standard model as a new
confining interaction is provided in
Appendix~\ref{sec:vectorconfinement}.

The most relevant degrees of freedom for hadron colliders are first,
the lightest colored particles, which will enjoy the largest
production cross sections, and second, vector octets, which can be
resonantly produced through mixing with the gluon.  Matter which is
charged under both QCD and the new confining group $G$ will typically
lead to a multiplet of colored pNGB's which will be among the lightest 
$G$-hadrons.  We will study here a pseudo-scalar octet of $G$-pions, $\pi_G^a$, which we will take to be
electroweak singlets. Pseudo-scalar octets will always have a minimal
pair-production cross-section at hadron colliders through their QCD
interactions. However, if a heavier spin-one $G$-hadron like a
$G$-vector meson $\rho^a_G$ is also present, the $\pi_G^a$
pair-production cross section can easily be enhanced by the potentially
large resonant production of the $\rho^a_G$ together with a large
coupling between $\rho^a_G$ and $\pi^a_G$.

We introduce here a phenomenological Lagrangian capturing the dynamics
of the pseudo-scalars $\pi_G^a$ together with an octet vector
$\rho_G^a$.  We will work with the effective Lagrangian
\begin{eqnarray}
-{\cal L}& =& - \frac{1}{2}\,D_\mu \pi^a_G \,D^\mu \pi_G^a + \frac{m_{\pi_G} ^ 2}{2} \pi^a_G \pi^a_G 
                           - \frac{1}{4}  \rho^{a\,\mu\nu}_G \rho^a_{G\,\mu\nu}+ \frac{m^2_{\rho_G}}{2} \rho^a_{G\,\mu}\rho_G ^{a\,\mu} 
 \nonumber \\                           
&&  +\frac{\tan \theta}{2}  \rho_G^{a\, \mu\nu} G^a_{\mu\nu} 
    + g_\rho f^{abc} \rho_G^{a\,\mu} \pi^b_G\,D_\mu \pi^c_G  \,,
                       \label{eq:leff}
\end{eqnarray}
together with two dimension-five operators allowing the $\pi^a_G$ to
decay, which will be discussed later.  This Lagrangian consists of mass and kinetic terms
for the $\rho_G^a$ and $\pi_G^a$, kinetic mixing between the
$\rho^a_G$ and QCD gluons $G^a$, and a $\rho_G^a$-$\pi_G^a$-$\pi_G^a$ vertex analogous to the familiar $\rho$-$\pi $-$\pi$ vertex in QCD.  Here the $\rho_G^a$ kinetic term
is written in terms of
\beq
\rho_G^{a\,\mu\nu} \equiv D^\mu\rho_G ^{a\,\nu}-D^\nu\rho_G ^{a\,\mu},
\eeq
with the covariant derivative $D^\mu\rho_G ^{a\,\nu}\equiv
\partial^\mu \rho_G^{a\,\nu} + i g_s f^{abc} G^{b\,\mu}
\rho_G^{c\,\nu}$. After making the field redefinition $G ^ a_\mu\to G
^ a_\mu + g_s \tan \theta \rho_{G\,\mu}^{a}$, the kinetic mixing
between the $\rho_G$ and the gluon is removed, while introducing a
coupling of the $\rho^a$ to quarks,
\beq
\label{eq:rhoqq}
-{\cal L}_{\rho_G q\bar q} = i g_s \tan \theta \,\rho^ {a}_{G\,\mu}\, \bar q\,t ^ a \gamma^\mu q .
\eeq
Here $t^a$ denotes the QCD generators. It is the coupling of Eq.~(\ref{eq:rhoqq}) to 
quarks which leads to resonant production of the $\rho_G$ at hadron colliders.\footnote{The field redefinition of the
gluon also shifts the value of $g_\rho$; we absorb this shift into
the definition of $g_\rho$.}

The theory described by Eq.~(\ref{eq:leff}) depends on four
parameters: the masses $m_{\rho_G}$ and $m_{\pi_G}$, the $\rho^a_G-G^a$ mixing
$\tan\theta$, and the $\rho^a_G-\pi^a_G$ coupling $g_\rho$.  The coupling
$\tan\theta$ governs the resonant $\rho_G^a$ production cross section,
while $g_\rho$ controls the relative branching fraction of the
$\rho_G$ into $G$-pions or back into $q\bar q$.  Although $g_\rho$
and $\tan\theta$ are independent parameters, for simplicity we
take them to be related according to
\beq
\label{eq:grhovstanth}
g_\rho = \frac{g_s}{\tan 2\theta} 
\,.
\eeq
This choice of relationship between $g_\rho$
and $\tan\theta$ is convenient for
comparison to a weakly coupled renormalizable coloron
model~\cite{Bai:2010dj}, which realizes similar phenomenology.  For
strong interaction models, one can estimate $g_\rho \sim
\sqrt{4\pi}$, which corresponds to a small value of $\theta \approx
0.14$. We will concentrate on this portion of parameter space, where
the dominant decay of the $\rho_G$ is to $G$-pion pairs. \footnote{For
$\theta\approx 0.14$, the
relation Eq.~(\ref{eq:grhovstanth}) yields a smaller value for
$\rho_G-G$ mixing than would be obtained from scaling the observed
$\rho-\gamma$ mixing in QCD \cite{Kilic:2008pm}.  The only importance
of this for our present purposes is in reducing the resonant $\rho_G$
cross-section relative to the QCD-like expectation, and our analysis
is in this sense conservative.} Once we have chosen a relationship
between $\tan\theta$ and $g_\rho$, the production times branching
ratio $\sigma (q\bar q\to \rho_G)\times \mbox{Br}(\rho_G\to q \bar q)
$ is fixed for a given $m_\rho$ and $\tan{\theta}$.  Results for other
choices of $g_\rho$ at a given $\tan\theta$ can be obtained by scaling
the branching ratio as desired, provided the total $\rho_G$ width
remains narrow.

The mass ratio $m_{\pi_G}/m_{\rho_G}$ is important for determining the
model's signatures at the LHC.  Previous LHC studies
\cite{Kilic:2008pm, Kilic:2009mi, Kilic:2008ub,  Kilic:2010et,
Sayre:2011ed} have focused on the region of parameter space where the {\it a priori} unknown ratio $m_{\pi_G}/m_{\rho_G}$ is chosen by scaling
from QCD, yielding~\cite{Kilic:2008pm}
\beq
\frac{m_{\pi_G}^ 2}{m_{\rho_G}^ 2} = 3 \left(\frac{\alpha_s}{\alpha}\right)  
             \frac{\left. \delta m_{\pi} ^ 2\right|_{EM}}{m_{\rho} ^ 2},
\eeq
where the observed electromagnetic contribution to the pion mass
splitting is $ \left.\delta m_{\pi} ^ 2\right|_{EM} \simeq \frac{3
\alpha}{4\pi} \, 2 \ln 2\, m_{\rho} ^ 2$ \cite{pionmass}.  This model for the unknown
$G$ dynamics yields $m_{\pi_G} \simeq 0.3 m_{\rho_G}$.
However, this
scaling relies critically on specific features of the QCD spectral
functions, whose genericity is unclear.  On general grounds, and 
avoiding the usual QCD-like $N_c$ scaling, we
may expect the $G$-pion mass to scale like
\beq
m^2_{\pi_G} \sim \frac{g_s ^ 2}{(4\pi) ^ 2} \Lambda_G^2,
\eeq
where $\Lambda_G$ is the scale where the $G$ interactions become strong.  For $\rho_G$ with mass
of order the cutoff, we can estimate
\beq
\frac{m_{\pi_G} }{m_{\rho_G}} \sim 0.1.
\eeq
Our main interest will be to establish the discovery reach in the
region of parameter space where $m_{\pi_G}$ is sufficiently small
compared to $ m_{\rho_G}$ that a flexible treatment of jets allows for
better separation of signal from background.  As we will see in
Section~\ref{sec:discovery}, our range of interest is therefore from
$m_{\pi_G}/m_{\rho_G} \approx 0.1$ up to the QCD-like value
$m_{\pi_G}/m_{\rho_G} \approx 0.3$, while for larger mass ratios
traditional multi-jet searches become more efficient.

Gauge invariance allows for additional renormalizable interactions
beyond those in the simplified model of Eq.~\er{leff} which we will
neglect, and which are further discussed in the Appendix~\ref{sec:vectorconfinement}.  Note that
the leading interactions of the new colored degrees of freedom with
the SM proceed only through QCD gauge interactions. This ensures
agreement with precision electroweak and flavor constraints.

The renormalizable interactions of Eq.~\er{leff} have a $Z_2$ symmetry
$\pi^a_G \rightarrow - \pi^a_G$ and do not yet allow the $\pi^a_G$ to
decay.  At dimension five, we can write down interactions which allow
either $\pi^a_G\to gg$ or $\pi^a_G \to q\bar q$.  Pion decay to gluons
is mediated by
\beq
\label{eq:Ogg}
{\cal O}_{\pi gg}=-\frac{g_s^2}{16 \pi^2 f_{\pi_G}} \Tr [t^a t^b t^c] \,\pi^a_G\, 
        \epsilon_{\mu\nu\rho\sigma} G^{b\,\mu\nu}G^{c\,\rho\sigma}.
\eeq
Here $f_{\pi_G}$ is the $G$-pion decay constant, $4\pi f_{\pi_G} \sim
\Lambda_G$. Using $\mbox{Tr}[t^a t^b t^c] =
\frac{1}{2}d^{abc} + \frac{i}{2}f^{abc}$, only the $d^{abc}$ part has
non-vanishing contributions. The operator ${\cal O}_{\pi gg}$ is
analogous to the operator mediating $\pi ^ 0\to\gamma\gamma$ in the
SM, and is naturally generated in many theories through triangular
anomaly diagrams, as we discuss further in the Appendix~\ref{sec:vectorconfinement}.  
In addition, depending on the details of the model, the pion may also decay to
quarks through the dimension-five operators
\beq
{\cal O}_{\pi q\bar q}=i\,\frac{c_{ij}^d}{M} \pi^a_G H \bar Q_L ^ i t ^ a\gamma ^ 5 d_R^j + 
     i\, \frac{c_{ij}^u}{M} \pi^a_G \tilde H \bar Q_L ^ i t ^ a\gamma ^ 5 u_R^j \,+\, h.c.
     \label{eq:pionqq}
\eeq
Here $M $ is an ultraviolet mass scale.  Assuming for simplicity that
these interactions are proportional to the SM Yukawa couplings,
$c_{ij}^{u,d} \propto Y_{ij}^{u,d}$, after electroweak symmetry
breaking these operators allow $G$-pion to decay through
\beq
\label{eq:Obb}
{\cal O}_{\pi q\bar q}=i\, \frac{m_q}{M} \pi^a_G \,\bar q \,t ^ a\gamma ^ 5 q  \,,
\eeq
where any order one coefficients have been absorbed into the
definition of $M$.  These couplings will favor $\pi^a_G\to b \bar
b$.\footnote{If the pion is sufficiently heavy that the decay to tops
is open, $m_{\pi_G}> 2m_t$, then $\pi^a_G\to\ttbar$ will dominate the
quark decay modes. The final state will be 4 $t$'s. We will largely be interested in $\pi^a_G$ below the $\ttbar $ threshold, and will consequently neglect the $\ttbar$
decay mode.}  At dimension-five level operators coupling $D_\mu
\pi^a_G$ to quarks also appear, ${\cal O} \propto D_\mu \pi^a_G \bar
q_{L,R} t ^ a \gamma^\mu \gamma ^ 5 q_{L,R}$.  The derivative portion
of these operators, upon use of the equations of motion, is equivalent
to Eq.~\er{Obb}, while the non-derivative portion contributes only to
3-body $G$-pion decays.

Since $G$-pion decay to both $b\bar b $ and to gluons proceeds through
higher dimension operators, details of the model can dramatically
affect the branching ratios of the $\pi^a_G$.  We will study discovery
prospects for either $\pi^a_G\to gg$ or $\pi^a_G\to b \bar b$ as the
dominant decay channel.  Strikingly, we will find that simple jet
substructure tools can dramatically enhance the prospects for
discovery even when the gluonic decays dominate.

The decay widths of the $\pi^a_G$ are given by
\beq
\Gamma (\pi_G\to gg) = \frac{5\alpha_s ^ 2}{192\pi ^ 3 f ^ 2_{\pi_G}} m_{\pi_G} ^ 3\,,
\eeq
through Eq.~\er{Ogg}, and
\beq
\Gamma (\pi_G\to b\bar b) = \frac{m_{\pi_G}}{16\pi}\left(\frac{m_b}{M}\right) ^ 2 
                \sqrt{1-4 m ^ 2_b/m ^ 2_{\pi_G}}\,,
\eeq
through Eq.~\er{Obb}.  The $G$-pions are narrow, and decay within the
detector for parameter choices of $f_{\pi_G}$ and $M$ in this paper.

\begin{figure}[]
\begin{center}
\includegraphics[width=0.48\textwidth]{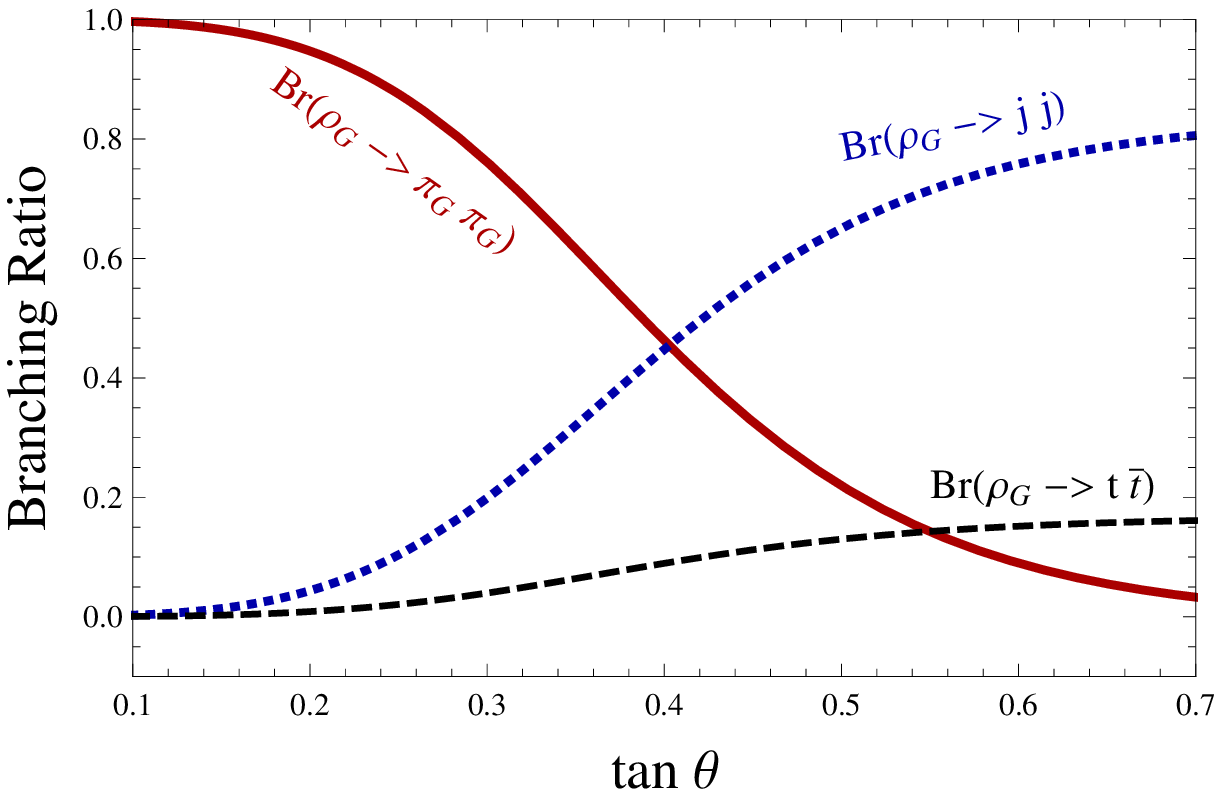} \hspace{2mm} 
\includegraphics[width=0.48\textwidth]{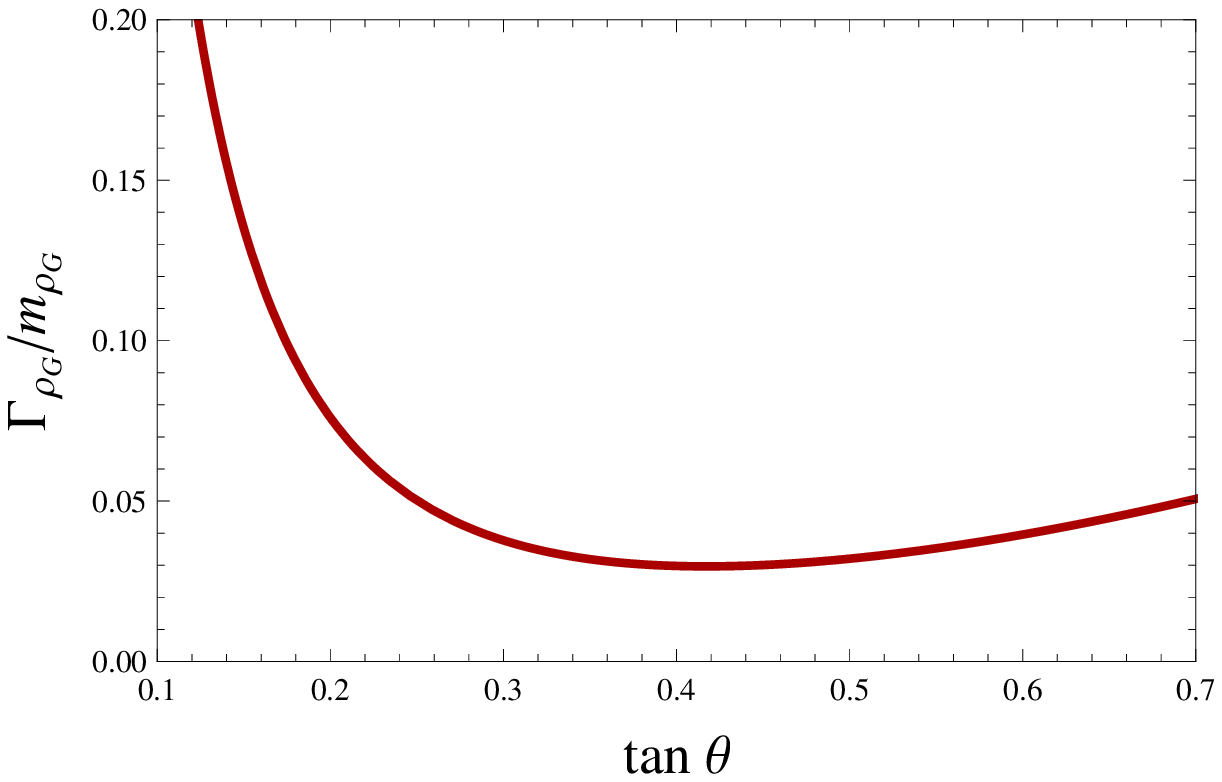} \hspace{2mm}
\caption{Left panel: the branching ratios of $\rho_G^\mu$ into different 
modes as a function of the mixing angle. Right panel: the $\rho_G^\mu$ 
width over its mass as a function of $\tan{\theta}$. The daughter particle masses are neglected.
}
\label{fig:branchingwidth}
\end{center}
\end{figure}

The decay widths of the $\rho^a_G$ into two quarks and into two
$G$-pions are given in terms of $\tan \theta$ as
\bea
\Gamma(\rho_G \rightarrow q\,\bar{q}) &=& \frac{\alpha_s}{6} \tan^2{\!\theta}\,m_{\rho_G}
\left( 1\,-\,\frac{4\,m_q^2}{m^2_{\rho_G}}\right)^{\! 1/2} \,, \nonumber \\
\Gamma(\rho_G \rightarrow \pi_G\,\pi_G) &=&  \frac{\alpha_s}{ 8\,\tan^2{2\theta}}\,m_{\rho_G}
\left( 1\,-\,\frac{4\,m_{\pi_G}^2}{m^2_{\rho_G}}\right)^{3/2}  \,,
\label{eq:decay}
\eea
where the branching fraction into quarks is per flavor. For a heavy
$\rho^a_G$ much above the $\pi^a_G$ and the top quark masses, we have
the branching ratios and the width over mass ratio shown in
Fig.~\ref{fig:branchingwidth}.

\begin{figure}[]
\begin{center}
\includegraphics[width=0.48\textwidth]{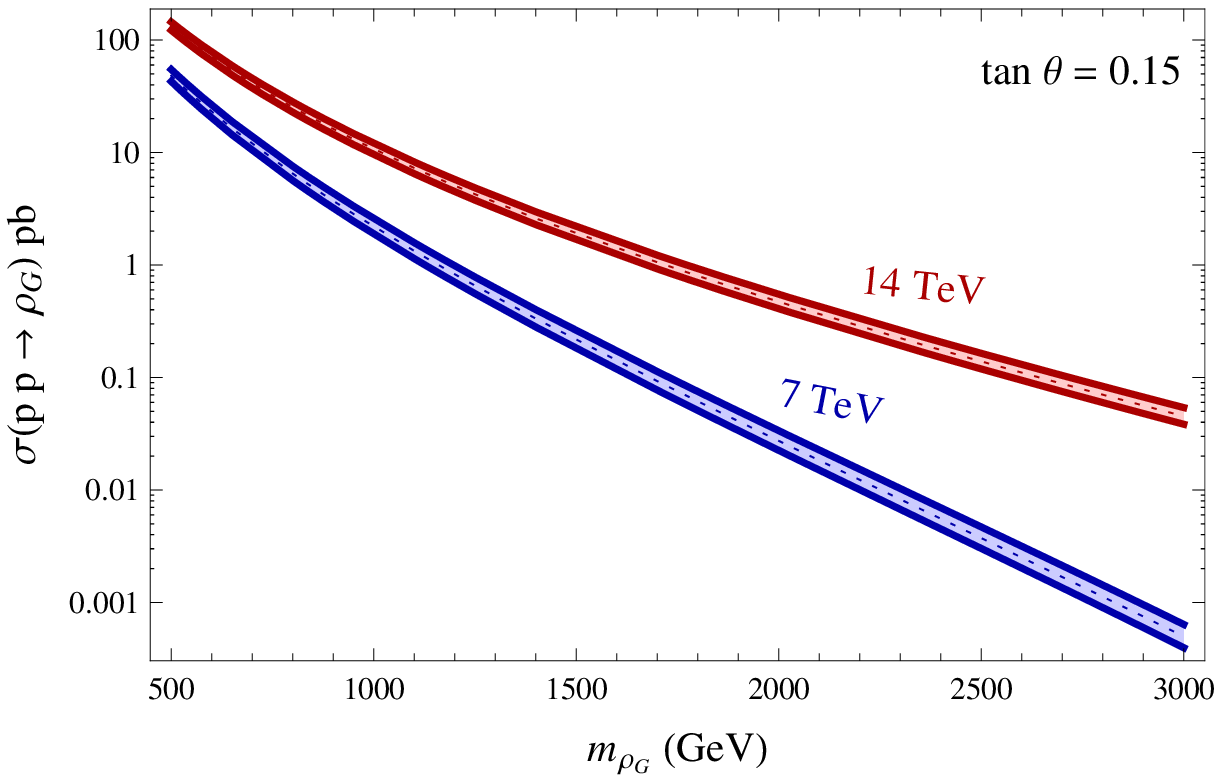} \hspace{2mm} 
\includegraphics[width=0.48\textwidth]{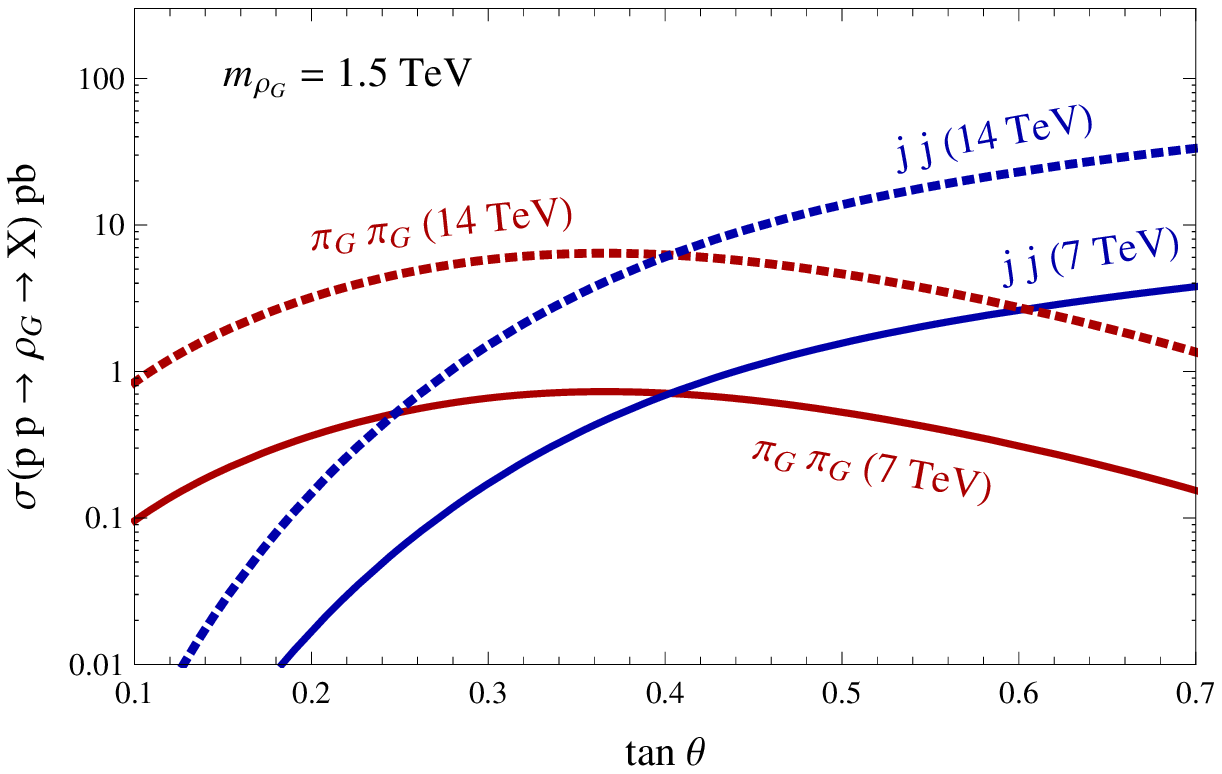} \hspace{2mm}
\caption{Left panel: the production cross section of $\rho_G$ at the
LHC for $\tan{\theta}=0.15$. The range of cross section is for two different renormalization
scales: $\frac{m_{\rho_G}}{2}$ (upper) and $2 m_{\rho_G}$ (lower).  Right
panel: the production cross section of different decaying modes, where
the renormalization scale is fixed to be $m_{\rho_G}$.  }
\label{fig:production}
\end{center}
\end{figure}

Using the narrow width approximation to estimate the cross section for
producing a $\rho^a_G$ in the $s$-channel gives
\beq
\sigma (q\bar{q} \to \rho_G^\mu ) \approx 
          \frac{ 8 \pi^2 \alpha_s \tan^2{\!\theta} }{9 m_{\rho_G}}  \; 
          \delta \left(\sqrt{\hat{s}} - m_{\rho_G}\right)  ~~.
\eeq
Convoluting this partonic cross section with the
MSTW~\cite{Martin:2009iq} PDFs yields the LHC production cross
sections shown in Fig.~\ref{fig:production}.

\begin{figure}[]
\begin{center}
\includegraphics[width=0.48\textwidth]{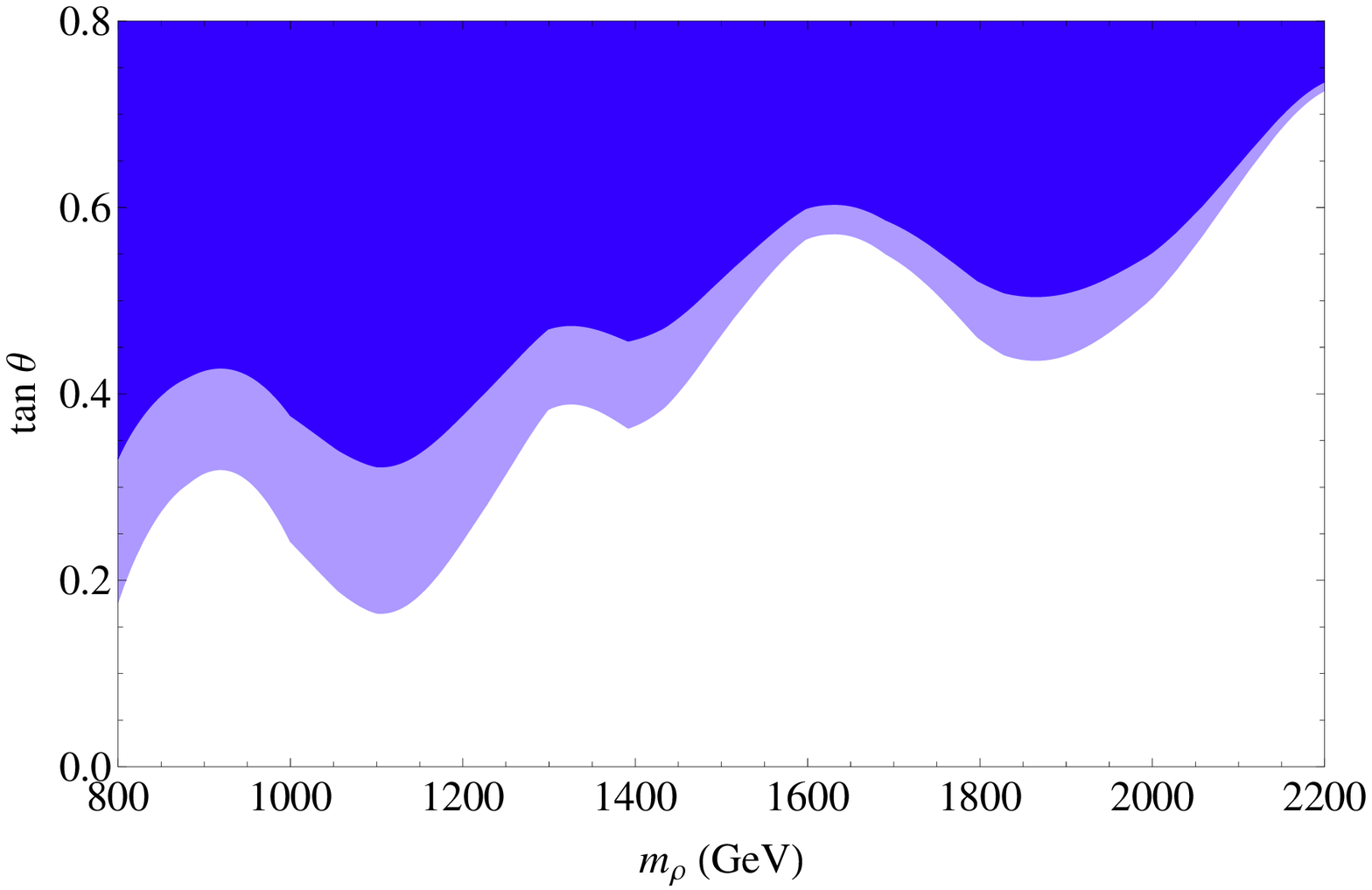} \hspace{3mm}
\includegraphics[width=0.48\textwidth]{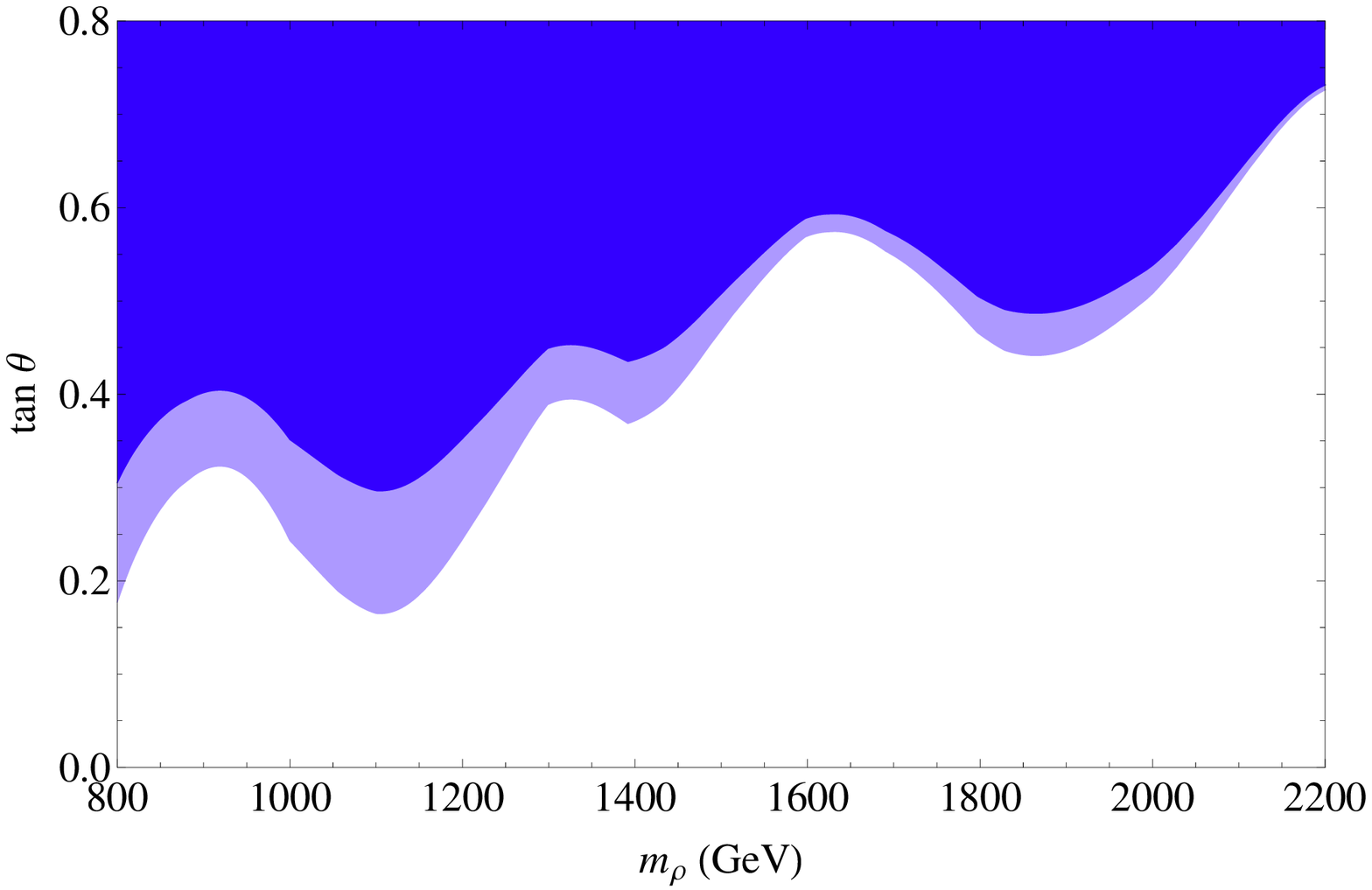}
\caption{Dijet constraints on the model parameters at 95\% C.L from
Atlas dijet narrow resonance searches with 163 pb$^{-1}$
luminosity~\cite{dijetupdate}. The shaded regions are excluded. The
two boundary lines are for (lower) 100\% efficiency to detect
$\pi_G^a$ as a single jet and (upper) 0\% efficiency to detect
$\pi_G^a$ as a single jet.  Branching ratios are given by
Eq.~(\ref{eq:decay}).  The left panel shows limits for
$m_{\pi_G}/m_{\rho_G} =0.1$ while the right panel shows limits for
$m_{\pi_G}/m_{\rho_G} = 0.3$.  }
\label{fig:dijetconstraints}
\end{center}
\end{figure}

The parameter space of $\tan{\theta}$ and $m_{\rho_G}$ is subject to
various constraints, most notably $t\bar t$ and dijet resonance
searches. The latest $t\bar t$ narrow resonance searches with 200
pb$^{-1}$ at Atlas~\cite{Atlasttbar} do not constrain our model
parameter space, because of the suppressed branching ratio of
$\rho_G\rightarrow t\bar t$ seen in Fig.~\ref{fig:branchingwidth}.
However, the dijet resonance searches \cite{dijetcrosssection, Aad:2011aj,dijetupdate} do
constrain our parameter space, and the limits are shown in
Fig.~\ref{fig:dijetconstraints}.
Applying the dijet limits to our model is not completely
straightforward, as there will be some nonzero efficiency for $\rho_G \to \pi_G \pi_G$ events to be reconstructed in the $\rho_G\to j
j$ sample.  This efficiency depends on the mass ratio
$m_{\pi_G}/m_{\rho_G}$ and on the jet algorithm used by Atlas. In
Fig.~\ref{fig:dijetconstraints}, we show the constraints in the
$m_{\rho_G}-\tan{\theta}$ plane from the dijet resonance search for
two fixed ratios of $m_{\pi_G}/m_{\rho_G}$.  The two boundaries in
each plot correspond to (lower) 100\% efficiency to detect $\pi_G^a$
as a single jet and (upper) 0\% efficiency to detect $\pi_G^a$ as a
single jet, with the branching ratios as given in
Eq.~(\ref{eq:decay}).  The limits are sensitive to the mass ratio only
through the factor $( 1\,-\,4\,m_{\pi_G}^2/m_{\rho_G} ^ 2) ^ {3/2} $,
and thus the limits are broadly similar for mass ratios $\ll 1$.

Finally, there are few meaningful limits on the $G$-pion masses.
While light $G$-pions can be pair-produced through $gg\to \pi_G\pi_G$
at the Tevatron, no multi-jet or multi-$b$ searches have limited their
masses \cite{Kilic:2008pm,Bai:2010dj}.  Introducing operators beyond
those in the simplified model Lagrangian of Eq.~(\ref{eq:leff}) can
potentially lead to indirect limits, as we discuss further in the
Appendix~\ref{sec:vectorconfinement}.

\section{Discovery potential}
\label{sec:discovery}

In this section we will detail discovery prospects for $\pi^a_G$ and
$\rho_G^a$.  The process we will use for discovery is $\pi^a_G$ pair
production from an initial $\rho^a_G$ resonance, $q\bar q\to\rho_G\to
\pi_G\pi_G $, followed by $\pi_G\to b\bar b, gg$.  This process, in
contrast to non-resonant $G$-pion pair production, is particularly
useful in theories with hierarchical spectra, where the $G$-pions
coming from the $\rho_G$ are sufficiently boosted that their daughter
partons have a reduced probability to be reconstructed as separate
jets. In this regime, the kinematics of the boosted $\pi^a_G$ allow
for easier separation of signal from background than do the
non-boosted $\pi^ a_G $ coming from nonresonant QCD pair production.
Simple jet substructure analyses then suffice to give excellent
discovery reach over much of the simplified model parameter space.

The main task is to distinguish a collimated perturbative two-body
decay $\pi^a_G\to j j$ from a QCD jet. When the pions are sufficiently
boosted that both their daughter jets have a moderate probability to
be contained in a standard ($R\geq 0.7$ anti-$k_T$ ~\cite{Cacciari:2008gp}) jet, the jet mass
alone can provide significant improvements over a standard dijet
search, as we will discuss below.  For slightly less boosted
$G$-pions, a more involved analysis improves the prospects.  We employ
a fat jet analysis based on the mass drop procedure pioneered by
\cite{Butterworth:2002tt, Butterworth:2008iy}.  Specifically, we cluster the events on a
large angular scale ($R=1.2$) using the Cambridge/Aachen ($\rm{C/A}$)
algorithm and require two fat jets with $p_T>p_{T,cut}$.  The fat jets
are then each iteratively decomposed by undoing the clustering
sequence step by step in search of a splitting which resembles a
perturbative decay. At each splitting of a parent $J$ to two daughters
$j_1,\,j_2$ with $m_{j_1} > m_{j_2}$, we check whether the splitting
\begin{itemize}
  \item shows a sudden drop in the jet mass, $m_{j_1} < \mu\,m_{J}$,
  \item and is relatively symmetric, ${\rm min}(p^2_{T\,j_1}, p^2_{T\,j_2}) \Delta
R^2_{j_1, j_2}/m^2_j > r_{xy}$.
\end{itemize}
Optimal values for the mass drop variable $\mu$ and the symmetric
splitting cut $r_{xy}$ will be chosen below.  If both conditions are
satisfied, one identifies $J$ as the fat jet and $j_{1, 2}$ as the
subjets and exits the loop. Otherwise, one replaces $J$ by $j_1$ and
repeats the previous procedure. In addition to vetoing QCD, the mass
drop analysis~\cite{Butterworth:2008iy} helps clean up the jets and
improves mass resolution.\footnote{We do not implement filtering \cite{Butterworth:2008iy} or
other jet grooming tools \cite{Ellis:2009su}, nor do we simulate pileup,
though in a full analysis both pileup and jet grooming will be necessary.}

Another observable which can distinguish a perturbative decay from a
QCD branching is the jet shape {\em $N$-subjettiness}
\cite{Thaler:2010tr}.  Given a jet found with initial radius $R$ and
a set of $N$ subjet centers $j_k$ found (with some algorithm) inside
the jet, the $N$-subjettiness of the jet is
\beq
\label{eq:Nsubj}
\tau_N = \frac{\sum_i p_{T,i} \min [\Delta R _{ik}]}{\sum_i p_{T,i} R},
\eeq
where the sum runs over the particles in the jet, and $\Delta R_{ik}$
is the distance between the $i^{\mathrm{th}}$ particle and the
$k^{\mathrm{th}}$ subjet axis.  Jets with smaller (larger) values of
$\tau_N$ have radiation more (less) concentrated around the subjet
axes, and are therefore more (less) amenable to a description in terms
of $N$ subjets. Since the QCD background tends to have larger values of $\tau_2/\tau_1$ than the signal, the ratio $\tau_2/\tau_1$ can be used as a good
discriminant to reduce QCD backgrounds.  We find that while
$N$-subjettiness and the mass drop procedure are clearly correlated,
they are sufficiently distinct that incorporating a cut on
$N$-subjettiness marginally improves discovery sensitivity.

We incorporate both the mass drop procedure and $N$-subjettiness into
a simple and flexible tagger designed to discriminate a boosted
$G$-pion from a QCD jet.  The tagger constructs a fat $\rm{C/A}$ jet with
$R=1.2$.  From the constituents of this fat jet, we construct two
exclusive subjets using the $k_T$ algorithm, yielding the two subjet
axes we use to evaluate $\tau_2/\tau_1$.  We require that the fat jet
passes a cut on $\tau_2/\tau_1$ in addition to the mass drop criterion.
Two (C/A) subjets $j_1$ and $j_2$ are identified in the mass drop
procedure, and the tagger incorporates cuts on both the sum of their
transverse momentum $p_{1\,T} +p_{2\,T}$ as well as their invariant mass
$m_{j_1 j_2} $. The specific values used for the cuts will be
discussed further below. The jet mass alone is a useful jet substructure 
variable \cite{jetmass}, and we will also demonstrate the reach
of a search which uses only the jet mass.

Before presenting results, we describe our simulation procedure.  The
production cross sections from Fig.~\ref{fig:production} vary from 100
fb to 400 fb at the 7 TeV LHC for the mixing angle $0.1 < \tan{\theta}
< 0.4$. To be concrete, we choose $\tan{\theta} = 0.15$ or
$\sigma(u\bar{u} \rightarrow \rho_G \rightarrow \pi_G\, \pi_G) \approx
200$~fb for $m_{\rho_G} = 1.5$~TeV throughout this section. The backgrounds are dominated by QCD dijets. We use the leading order cross-section as calculated in
MadGraph~\cite{Alwall:2007st}, as comparison with measured dijet cross-sections
\cite{dijetcrosssection} indicates good agreement (i.e., $K$-factors
near unity) in the high-$p_T$, large invariant mass regime of
interest.  Renormalization and factorization scales are set at
$\mu=m_{\rho_G}$.  The subleading $W+$ jets and $t\bar{t}$ backgrounds
are negligible compared to the dijet background.  Both signal and
background events are generated with MadGraph~\cite{Alwall:2007st}
using CTEQ6L PDFs \cite{Pumplin:2002vw} and showered in Pythia
6.4.24~\cite{Sjostrand:2006za}. We then bin visible particles with
$|\eta | < 2.5$ into massless $0.1\times 0.1$ calorimeter cells and pass
to FastJet~\cite{fastjet,Cacciari:2005hq} for clustering and
subsequent jet analysis.

Recent studies have demonstrated that Pythia and Herwig show
reasonable agreement both with each other and with the data for jet
masses in the range of interest.  At high masses, Pythia and Herwig
give nearly indistinguishable predictions for large C/A jets put
through the mass drop procedure. Results for $R=1.0$ anti-$k_T$ jets
indicate that while overall agreement is good, Pythia tends to
underpredict QCD jet masses by 15-20\% in the mass range $100~\gev <
m_j<200~\gev$~\cite{Atlasjets}.  We thus conclude that the numbers we
will obtain for the analysis built on the full tagger are
representative, while the alternate analysis using only anti-$k_T$ jet
masses is likely to be slightly optimistic due to the tendency of
Pythia to underpredict background QCD jets in the mass range of
interest.  The performance of the $G$-pion tagger on QCD dijets can be
validated using dijet events where only one jet has a mass within the
$G$-pion mass range and the other is light ($m_j \ltap 50$~GeV).

\subsection{$\pi_G \to  gg$}
\label{sec:fourgluon}
For the case where $\pi_G$ dominantly decays into two gluons, the
signal is $q\bar{q} \rightarrow \rho_G \rightarrow \pi_G\, \pi_G
\rightarrow 4\,g$'s. We will first illustrate our reconstruction
procedures at the specific point $m_{\rho_G} = 1.5$~TeV and $m_{\pi_G}
= 300$~GeV, and then present the discovery potential for other
combinations of $m_{\rho_G}$ and $m_{\pi_G}$.

\begin{figure}[]
\begin{center}
\includegraphics[width=0.48\textwidth]{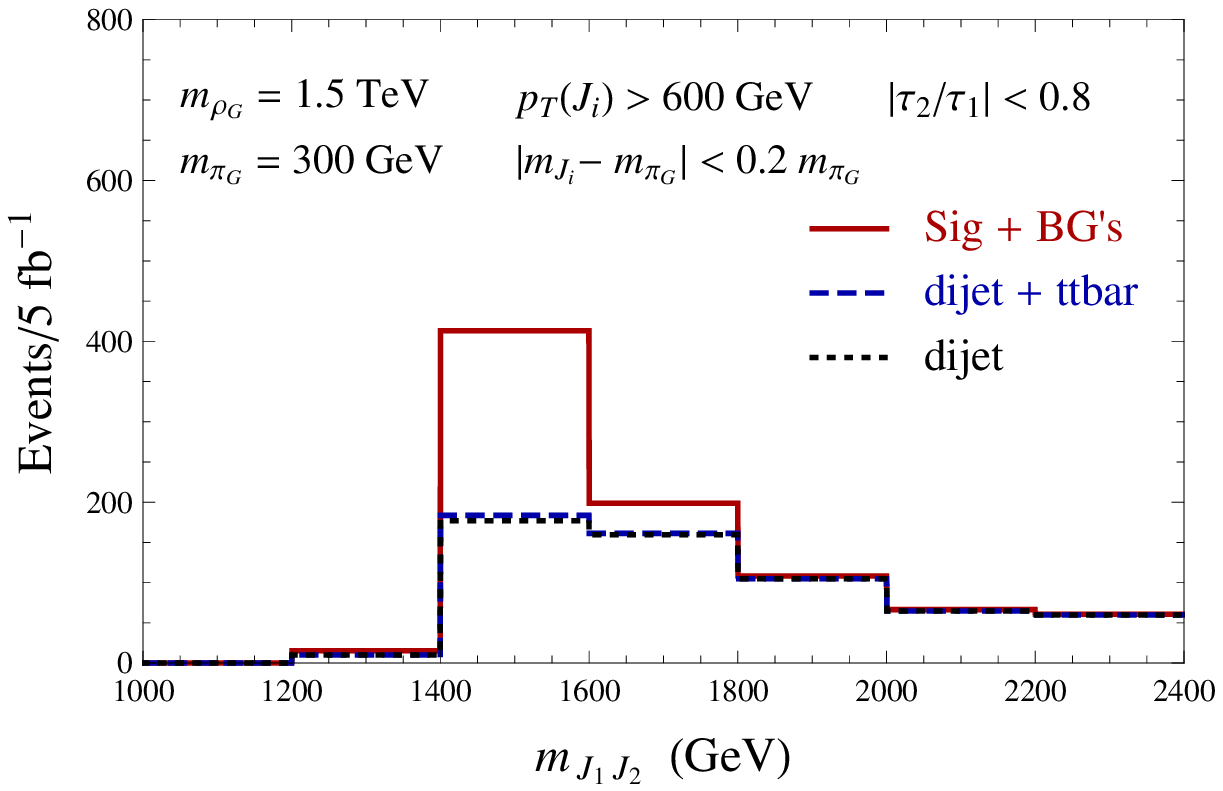} \hspace{3mm} 
\includegraphics[width=0.48\textwidth]{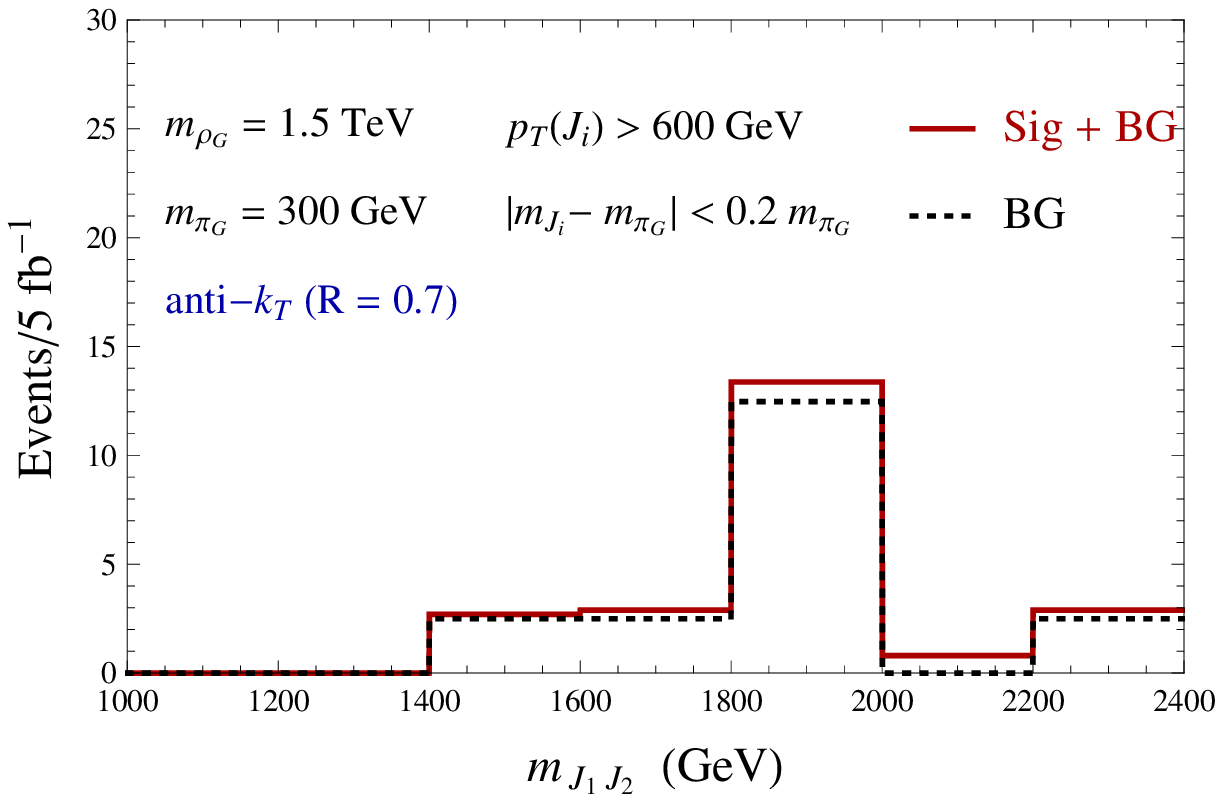}
\caption{Left panel: the numbers of events of signal and backgrounds at the 
7 TeV LHC after the jet substructure analysis. Right panel: the same as the 
left panel but for the analysis without using jet substructure.}
\label{fig:PmassGp1500}
\end{center}
\end{figure}

We find that the final discovery significance is relatively
insensitive to varying $R$, the mass drop $\mu$, and the
symmetricity cut $r_{xy}$. We fix these parameters at $R =1.2$, $\mu
= 0.3$ and $r_{xy} = 0.3$ to generate the left panel in
Fig.~\ref{fig:PmassGp1500}.
After implementing the substructure tagger, we require two tagged
$G$-pion candidates in the event with $p_T(j_i) > 600$~GeV.  Note the
$p_T$ cut is placed on the final $G$-pion candidate rather than the
initial fat jet.  Taking into account the jet energy
resolution~\cite{Atlas:2010bc} and jet mass resolution \cite{Atlasjets}, we
further require additional mass window cuts $|m_{J_i} - m_{\pi_G}| <
0.2\, m_{\pi_G}$. We show the histogram distributions of signal and
background events in the left panel of Fig.~\ref{fig:PmassGp1500}. As
demonstrated in this figure, the $t\bar{t}$ background only
contributes a tiny fraction of the total background; $W+j$ (not
shown) is below $\ttbar$.  Further imposing a mass window cut $|m_{J_1
J_2} - m_{\rho_G}| < 0.2\,m_{\rho_G}$, we find that the discovery
significance is $S/\sqrt{B}\approx 14$ for the 7 TeV LHC with 5
fb$^{-1}$ luminosity.

As a comparison, we also estimate the discovery significance obtained
by using a simple extension of the traditional dijet resonance
searches, performed with anti-$k_T$ jets at a 
fixed $R$.  On top of the usual cuts, namely jet $p_T$ cuts and the
dijet mass window cut, we also require both jet masses to be within
the $\pi_G^a$ mass window. The efficacy of this search depends on the
efficiency for a boosted $G$-pion to be contained within a single jet.
We show results for $R=0.7$, which is the largest standard cone size
in use at the LHC. Choosing a smaller value of $R$ will make this
simple search worse, while a value of $R=1.0$ as studied in
\cite{Atlas-topreco, Atlasjets} will improve the reach.  We show the
histograms of the signal and background events in the right panel of
Fig.~\ref{fig:PmassGp1500}. The discovery significance is around
$2\,\sigma$ for this parameter point, much poorer than the result obtained from the jet
substructure analysis. Because $R=0.7 < 2 m_{\pi_G}/p_{T, cut}$, the
jet clustering algorithm in the simple dijet search will typically not
capture all the signal decay products in a single jet and hence
suffers a reduction in the discovery significance. We have checked
that for a different mass combination, $m_{\rho_G} = 1.5$~TeV and
$m_{\pi_G} = 150$~GeV, the dijet resonance search supplemented with
jet mass can obtain a discovery sensitivity as good as the jet
substructure analysis.

\begin{figure}[t!]
\begin{center}
\includegraphics[width=0.46\textwidth]{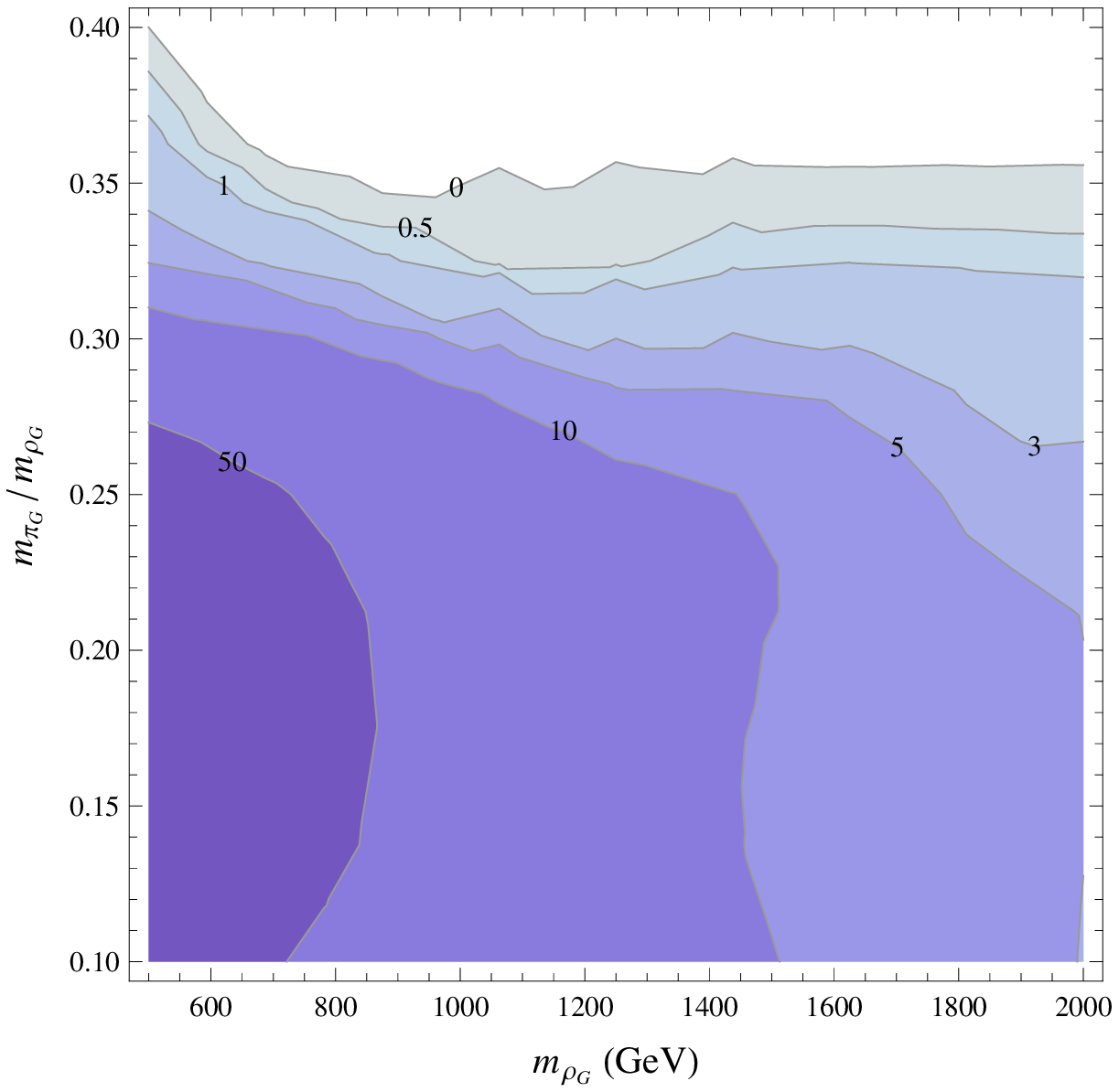} \hspace{2mm} 
\includegraphics[width=0.46\textwidth]{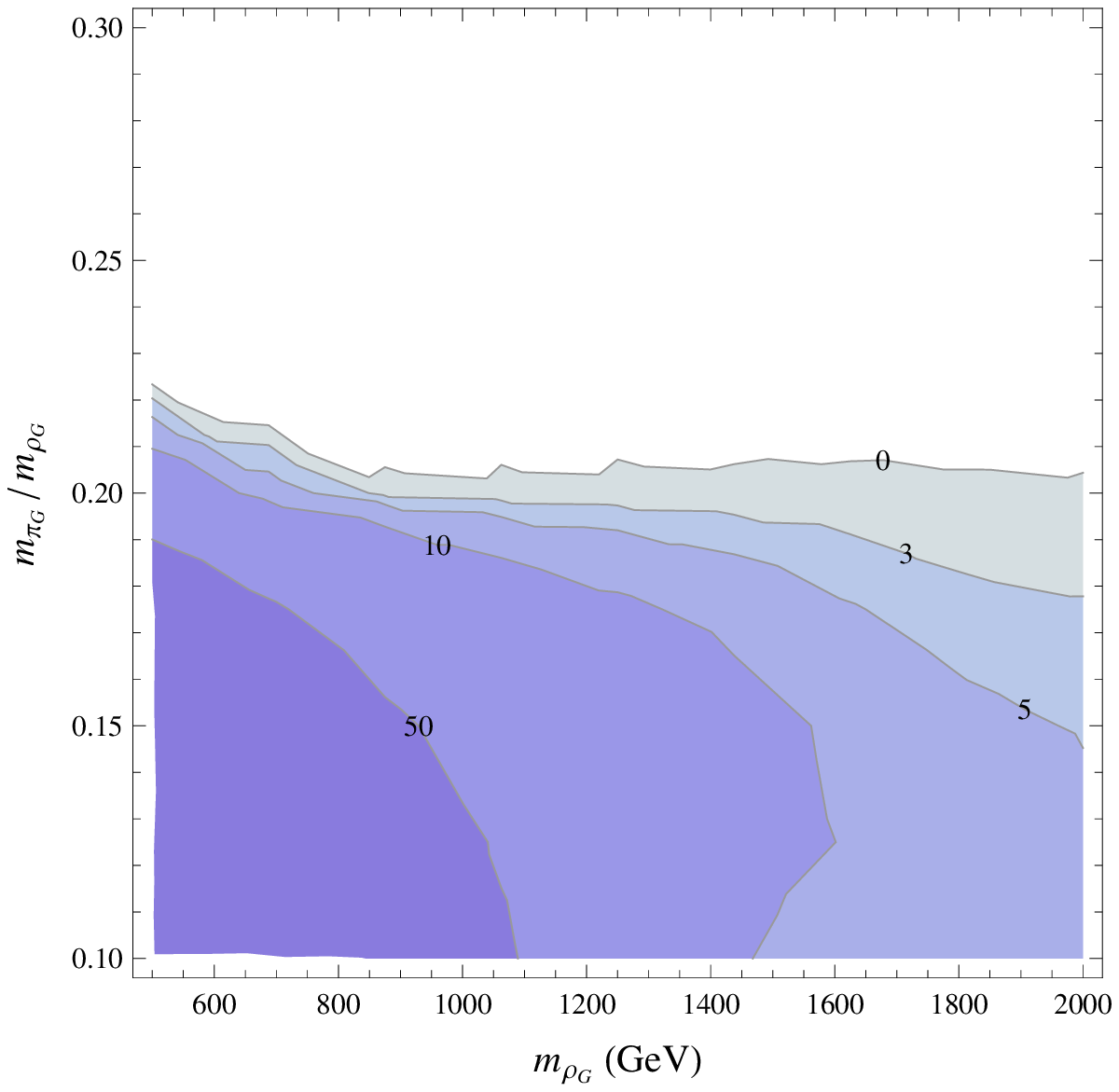} \hspace{2mm} 
\caption{Left panel: the discovery significance for different masses
of $\rho_G$ and $\pi_G$ for $\pi_G \rightarrow gg$. We scanned four
variables to find the optimized significance: the mass drop variable
$\mu$, the symmetric splitting cut $r_{xy}$, the $p_T$ cut of the fat
jets, and the $N$-subjettiness variable $\tau_2/\tau_1$. The numbers
besides each contour line are the significance in $\sigma$.  Right
panel: the same as the left panel but instead of using sophisticated
jet substructure analysis, only the jet masses are used in this plot.
}
\label{fig:fourgluon}
\end{center}
\end{figure}

For different mass combinations and especially when there are few
signal and background events, we use the Poisson distribution to
quantify the discovery significance as
\beq
\mbox{significance} \equiv \sqrt{-2 \ln{[e^{-S-B} (S+B)^B/\Gamma(B+1)}]}\,. 
\eeq
For different $\rho_G^a$ masses and different values of the mass ratio
$m_{\pi_G}/m_{\rho_G}$, we find the best discovery significance for
each mass point in the left panel~\footnote{Strictly, there is an
additional trials factor associated with the substructure searches
due to the unknown $m_{\pi_G}$.} shown in Fig.~\ref{fig:fourgluon} by
scanning the cut on $\mu$ from 0.2 to 0.4 with a step of 0.05, the cut
on $p_T$ from $m_{\rho_G}/3$ to $m_{\rho_G}/3 + 300$~GeV with a step
of 50 GeV, the cut on $r_{xy}$ from 0.2 to 0.4 with a step of 0.05,
the cut on $r_{xy}$ from 0.2 to 0.9 with a step of 0.1.  We further
require the mass window cuts $|m_{J_i} - m_{\pi_G}| < 0.2\, m_{\pi_G}$
and $|m_{J_1 J_2} - m_{\rho_G}| < 0.2\,m_{\rho_G}$.
To obtain the left panel of Fig.~\ref{fig:fourgluon}, we have scanned
7 different $\rho_G$ masses from 500 GeV to 2 TeV with a 250 GeV
interval and 7 different mass ratios from 0.1 to 0.4 with a 0.05
interval. As can be seen from Fig.~\ref{fig:fourgluon}, the jet
substructure analysis can discover the composite color octets for a
wide range of masses.  For smaller mass ratios of
$m_{\pi_G}/m_{\rho_G}$, the discovery significances are
better. Generically for $m_{\pi_G}/m_{\rho_G} > 0.3$, the jet
substructure analysis loses its effectiveness and one should instead
carry out a more traditional multi-jet resonance analysis to cover
this region~\cite{Kilic:2008pm, Dobrescu:2007yp,  Bai:2010dj, Plehn:2008ae}.

As a comparison, we show the discovery limit  in the right panel of
Fig.~\ref{fig:fourgluon} by using the ordinary dijet searches (with
anti-$k_T$ and $R =0.7$) and requiring the
two jet masses satisfying the mass window cuts $|m_{J_i} - m_{\pi_G}|
< 0.2\, m_{\pi_G}$ and $|m_{J_1 J_2} - m_{\rho_G}| <
0.2\,m_{\rho_G}$. From this plot, one can see that this very simple
analysis can discover the $\rho_G^a$ together with the $\pi_G^a$ especially for
$m_{\pi_G}/m_{\rho_G} < 0.2$. Comparing it with the left panel of this
figure, one can see that for the light $\rho_G^a$ mass region the
traditional dijet resonance searches with jet mass constraints are
even better than the complicated jet-substructure analysis.

\begin{figure}[]
\begin{center}
\includegraphics[width=0.6\textwidth]{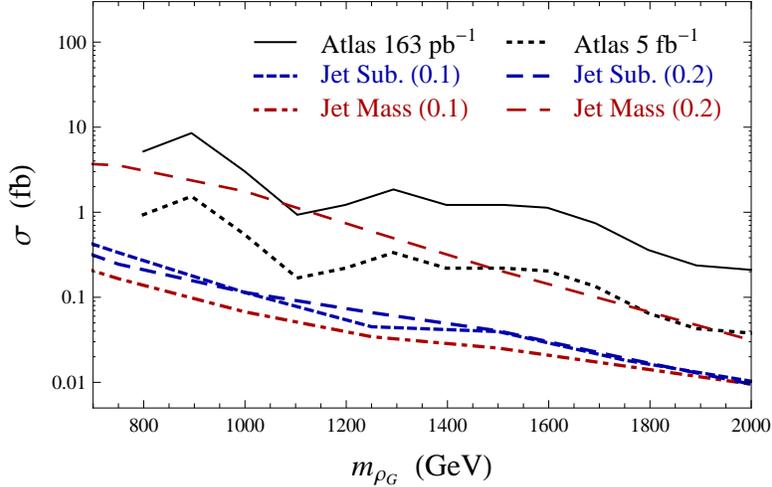} 
\caption{The 95\% C.L. exclusion limit on the resonance production
cross sections times branching ratio from different searches.  The
dotted black line is the projected Atlas exclusion limit at 5
fb$^{-1}$ based on the current limit with 163 pb$^{-1}$
luminosity~\cite{dijetupdate}. The numbers in parentheses denote the
ratio $m_{\pi_G}/m_{\rho_G}$.  }
\label{fig:comparison}
\end{center}
\end{figure}

Finally, we compare the sensitivities from the traditional dijet
searches and from the jet substructure searches in
Fig.~\ref{fig:comparison}.  We take the current results from narrow
resonance searches in dijets at Atlas with 163
pb$^{-1}$~\cite{dijetupdate}, and plot the projected 95\% C.L.
exclusion limit on the production cross section times dijet branching
ratio at 5 fb$^{-1}$ by assuming statistically dominated errors for
the backgrounds. We show results from the full jet substructure
analysis as well as the simple jet mass analysis (with $R =0.7$
anti-$k_T$) at the 7 TeV LHC with 5 fb$^{-1}$: here the vertical
axis is cross-section times $G$-pion branching ratio. As can be seen
from Fig.~\ref{fig:comparison}, for a small mass ratio
$m_{\pi_G}/m_{\rho_G}=0.1$ the simple jet mass analysis provides the
best exclusion limit, while for a small ratio
$m_{\pi_G}/m_{\rho_G}=0.2$ the full jet substructure analysis is the
most sensitive one. To produce this plot, we have neglected the
acceptance of the traditional dijet analysis, which is large and close
to $70\%\sim 80\%$. 
%

\subsection{$\pi_G \to b\bar b$}
\label{sec:fourb}

For the case where the main decay channel of $\pi_G$ is two $b$-jets,
we repeat the same analysis as the four gluon case except that we now
additionally demand two $b$ tags in the final state.  Although the
signal contains four $b$-quarks, we have found that requiring two
$b$-tags for the four daughter jets is sufficient to reject the
backgrounds.

The backgrounds now come from both two light jets with a double $b$
mistag, and two $b$-jets. After taking into account the $b$-tagging
efficiency, these two contributions to the background are
comparable. We assume a $b$-tagging efficiency of 60\% and a
mistagging efficiency of 2\% for light jets (the $c$-jet has a larger
mistagging efficiency which we compensate for by choosing a larger
value of mis-tagging efficiency for all light jets).  Improved
$b$-tagging efficiencies (70\% efficiency without increasing
mistagging rates) may be possible~\cite{AtlasB}, but as our final
state contains more hadronic activity than the $(0,1,2)\ell + 1$ fat
jet states where these studies were performed, we conservatively do
not use these improved numbers.

\begin{figure}[]
\begin{center}
\includegraphics[width=0.46\textwidth]{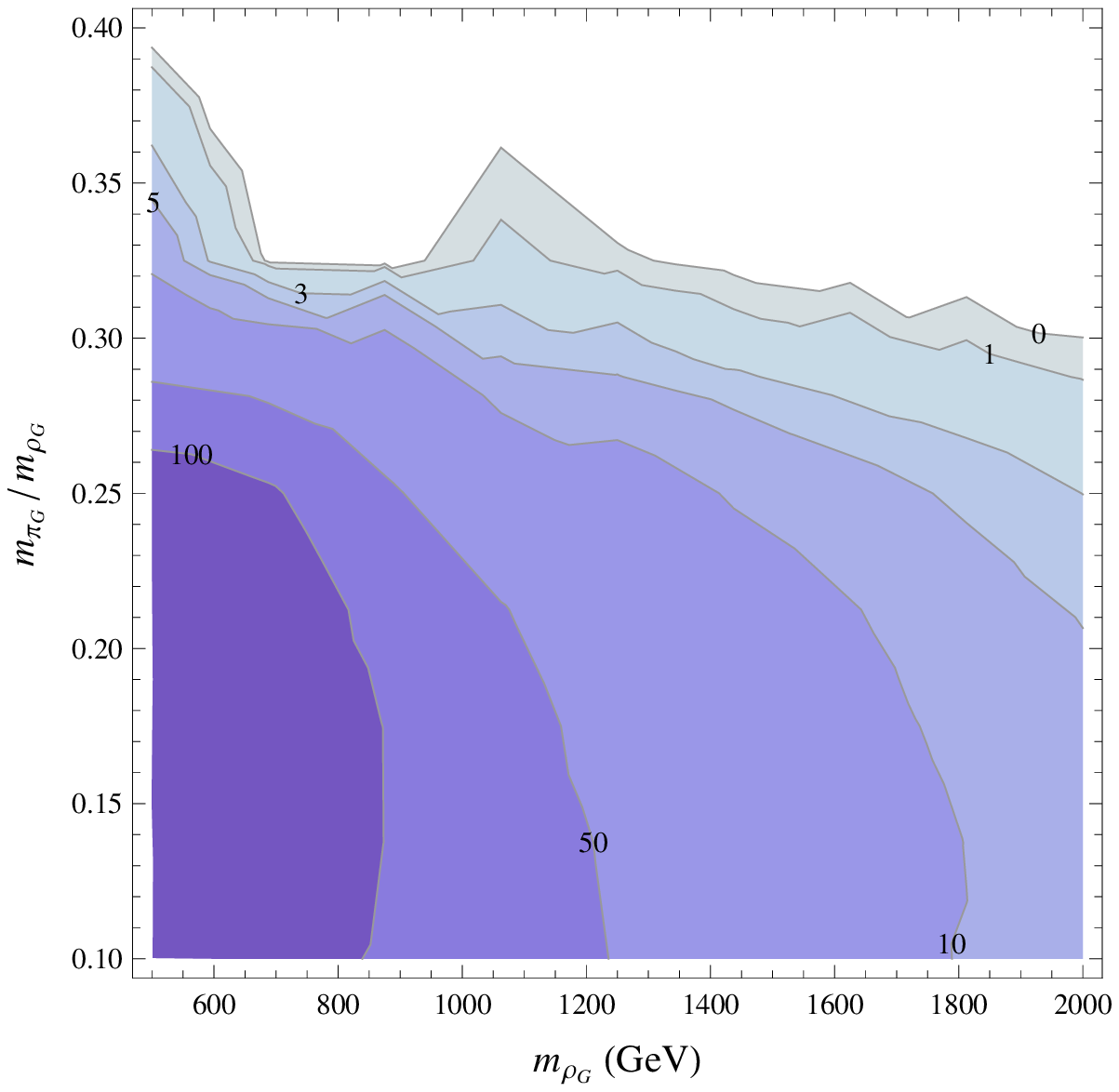} \hspace{2mm} 
\includegraphics[width=0.46\textwidth]{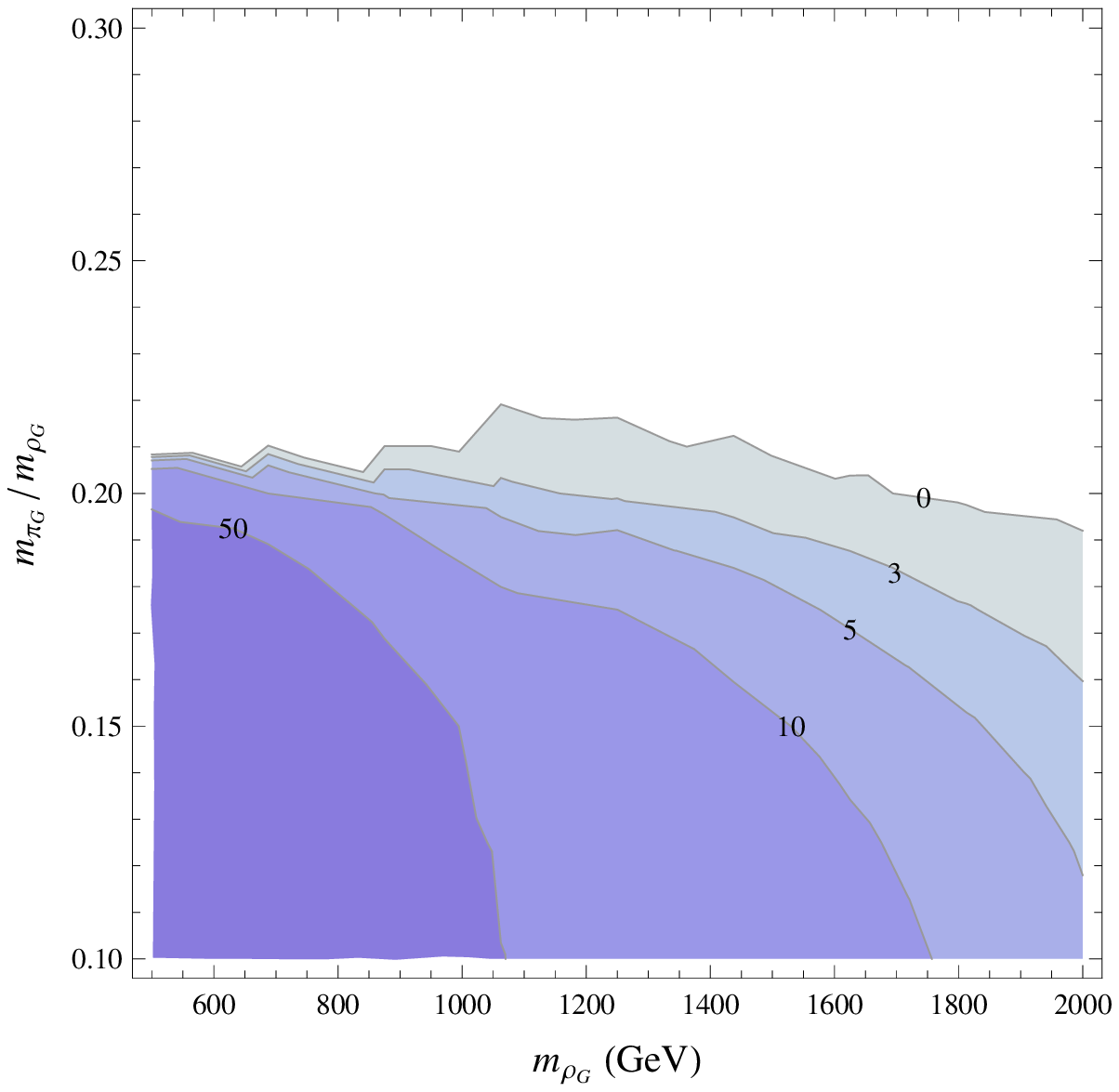} \hspace{2mm} 
\caption{The same as Fig.~\ref{fig:fourgluon} for the discovery
significance for different masses of $\rho_G$ and $\pi_G$ but for $\pi_G
\rightarrow b \bar b$.}
\label{fig:fourb}
\end{center}
\end{figure}

We require each fat jet to contain at least one $b$-tagged
subjet. Performing the same scan of mass combinations as in the four
gluon case, we find the discovery significance shown in
Fig.~\ref{fig:fourb}.
%

\section{Discussion and conclusions}
\label{sec:conclusions}

We have demonstrated the excellent potential of the 7 TeV LHC to
discover composite octets. The typically large branching fractions of
colored vector resonances to BSM daughters instead of to dijet final states 
makes their discovery difficult: the clean dijet signature has a suppressed rate,
while the multijet signature arising from $\rho_G\to\pi_G\pi_G$ can be difficult
to reconstruct.  We have demonstrated how jet substructure techniques 
improve the reconstruction of the $\rho_G$ and extend the discovery 
reach of the traditional dijet analysis for colored spin-1
resonances.  The topology of the final state in resonant $\rho_G$
production depends strongly on the mass ratio $m_{\pi_G}/m_{\rho_G}$. 
For large hierarchies,
$m_{\pi_G}/m_{\rho_G}\ltap 0.2$, a simple search augmenting the dijet
resonance search with an additional cut on jet mass works very well.  
For intermediate hierarchies, including the QCD-like region
where $m_{\pi_G}/m_{\rho_G}\approx 0.3$, a more involved jet substructure
analysis using a simple $G$-pion tagger gives the best sensitivity. 

The mass drop and $N$-subjettiness cuts used in the full $G$-pion
tagger are very effective at separating the perturbative decays from
QCD background.  On the other hand, they strongly shape the angular
distributions of the jets coming from $\pi_G ^ a \to jj$, which makes
probing the $G$-pion quantum numbers more challenging.  Requiring
that the jets coming from $\pi_G ^ a \to jj$ be sufficiently hard and
symmetric to be distinguished from typical QCD branchings selects only
the portion of the angular distribution which is transverse to the
axis of the $G$-pion boost.  This surviving slice of the angular
distribution contains minimal information and renders determination of
the $G$-pion spin difficult \cite{Englert:2010ud}.  In this regard,
the simple jet mass analysis offers some advantages, in the
highly boosted region of small $m_{\pi_G}/m_{\rho_G} $ where the jet mass search is
effective.  In this portion of parameter space, where hardness cuts on subjets are not critical for discovery,
the subjets identified within the boosted $G$-pion jet (for example,
by simply resolving at a small angular scale $R=0.3$ and selecting the hardest
subjet) preserve more of the underlying angular distribution and give 
good prospects for determining the $G$-pion Lorentz quantum numbers.

\begin{figure}[]
\begin{center}
\includegraphics[width=0.5\textwidth]{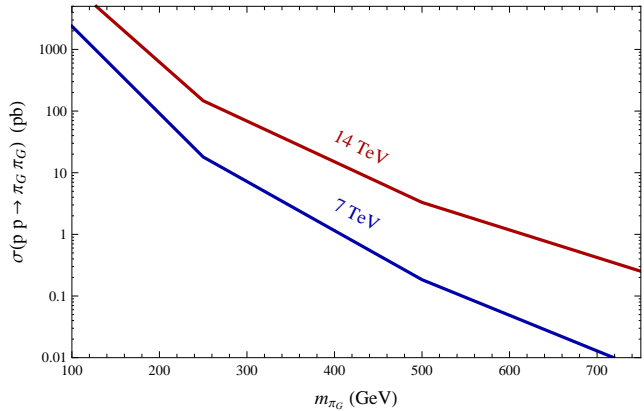} \hspace{2mm} 
\caption{The pair production cross section of two $\pi_G$'s from QCD interaction.}
\label{fig:QCDprod}
\end{center}
\end{figure}

We have focused on resonant $\rho_G\to \pi_G\pi_G$ production, neglecting the nonresonant QCD
pair-production of $\pi_G$. For completeness, we show the production
cross section of $p p \rightarrow \pi_G \pi_G$ in
Fig.~\ref{fig:QCDprod} for the 7 TeV as well as 14 TeV LHC.
Nonresonant pair production can allow octet $G$-pions to be discovered
at the LHC using mass window cuts~\cite{Kilic:2008ub, Dobrescu:2007yp}.  
We want to emphasize that the studies performed in this paper could simultaneously lead to the discovery of
{\it two} composite color octet particles.  Alternatively, if the octet
$G$-pion is first found in non-boosted multi-jet final states, its
mass may be used as an input to the $G$-pion tagger in a search for
the $\rho_G$.  The main advantage of the techniques presented here is the
improved sensitivity to the $\rho_G^a$.
Especially for theories with $m_{\pi_G}/m_{\rho_G} \ltap 0.2$,
the jet substructure analyses we propose could be the unique way to
discover the $\rho^a_G$, and to understand the detailed properties of a new
strong interaction.

\bigskip
{\em Acknowledgements:} Thanks to W.~Goldberger, S.~Hoeche, and D.~Zerwas for useful discussions.  We are grateful to
M.~Cacciari and G.~Salam for assistance with Fastjet and Peter Skands
for assistance with Pythia. This material is based upon work supported 
in part by the National Science Foundation under Grant No. 1066293 and 
the hospitality of the Aspen Center for Physics.  JS was supported in 
part by DOE grant DE-FG02-92ER40704.  This work was supported in part 
by the facilities and staff of the Yale High Performance Computing
Center, and by the NSF grant CNS 08-21132. SLAC is operated by
Stanford University for the US Department of Energy under contract
DE-AC02-76SF00515.

\appendix
\section{Vector-like confinement}
\label{sec:vectorconfinement}

In this appendix we show how a well-motivated extension of the SM maps
on to the simplified model discussed in Section~\ref{sec:model}.

We suppose here that the new gauge sector includes a fermion species
$\Psi$ which transforms as a fundamental under QCD and as $G_\psi$
under $G$,
\beq
\Psi_L = (G_\psi, 3), \phantom{spacer} \overline{\Psi}_R = (\overline{G}_\psi, \overline{3}).
\eeq
The theory possesses a global $\Psi $ flavor symmetry $SU(3)_L\times
SU(3)_R$ other than the global baryon symmetry in the $G$-sector.
When the gauge group $G$ confines at a scale $\Lambda_G$, this chiral
symmetry is broken down to the weakly gauged diagonal $SU(3)_c$,
leaving an octet of pNGB's which we denote $\pi_G^a$. In the strong
interacting $G$-sector, there could also exist vector mesons as well
as axivector mesons. As the lightest axivector meson is in
principle heavier than the vector meson as is the case in QCD, and 
has suppressed resonant cross sections, we only
consider the vector meson in the following.

We reproduce here the minimal Lagrangian of Eq.~\er{leff}, 
\begin{eqnarray}
-{\cal L}& =& - \frac{1}{2}\,D_\mu \pi^a_G \,D^\mu \pi_G^a + \frac{m_{\pi_G} ^ 2}{2} \pi^a_G \pi^a_G 
                           - \frac{1}{4}  \rho^{a\,\mu\nu}_G \rho^a_{G\,\mu\nu}+ \frac{m^2_{\rho_G}}{2} \rho^a_{G\,\mu}\rho_G ^{a\,\mu} 
 \nonumber \\                           
&&  +\frac{\tan \theta}{2}  \rho_G^{a\, \mu\nu} G^a_{\mu\nu} 
    + g_\rho f^{abc} \rho_G^{a\,\mu} \pi^b_G\,D_\mu \pi^c_G  \,,
                    \label{eq:leff2}
\end{eqnarray}
and comment on additional possible terms and their consequences.

First, terms polynomial in $\pi_G^a$ will generically be present, but
suppressed due to the approximate shift symmetry of the $G$-pions.
The cubic interaction $\mu\,d^{abc} \pi_G ^ a\pi_G ^ b\pi_G ^ c$ is
notable as it breaks parity; we set this term to zero.  There is also
a coupling between the $G$-pions and the SM Higgs, $\lambda_{\pi H}\,
(\pi^a_G \pi^a_G) |H|^2$, which can lead to indirect limits on $m_\pi$
through its effect on Higgs production through gluon fusion
\cite{HiggsLimits}.  In our scenario, the Higgs is not part of the
confining gauge sector, and hence $\lambda_{\pi H}$ is radiatively
generated.  With $\lambda_{\pi H}\ll 1$, the $\pi^a_G$ may safely have
masses in the $100-200~\gev$ range. The pion number symmetry in the
$G$-sector is broken by the anomalous coupling among $\pi^a_G$ and two
gluons in Eq.~(\ref{eq:Ogg})~\cite{Bai:2010mn,
Freitas:2010ht}. Additional higher-dimensional operators can directly
couple $\pi_G$ to SM quarks and mediate $\pi_G$ decaying into quarks
as shown in Eq.~(\ref{eq:pionqq}).

It is also possible to write additional interactions for the $\rho_G$.
The renormalizable interaction
\beq
 {\cal O}_{\rho \rho g}= \lambda_{\rho \rho g} f^{abc} G_{\mu \nu}^a \rho^{b\,\mu}_G\rho^{c\,\nu}_G\,,
\eeq
contributes to $\rho_G$ pair production. There are two more
renomalizable operators containing only the $\rho_G$ field and we
neglect them here.  At dimension-6 level, we find the operator
\beq
{\cal O}_{\rho gg} =  \frac{i \lambda_{\rho gg}}{4 \pi \Lambda_G^ 2} f^{abc} 
                             \rho^{a\mu}_{G\,\nu} G^{b\nu}_{\rho} G^{c\rho}_{\mu},
\eeq
which could be the leading contribution to resonant $gg\to\rho_G$
production \cite{Chivukula:2001gv}.  As the gluon-gluon luminosity at
the LHC is large, this operator can have a noticeable impact on the
resonant $\rho_G$ cross-section despite its high
dimension~\cite{Kilic:2008ub}.  We conservatively do not include this
process when we evaluate $\sigma (pp\to\rho_G)$. Gauge invariance also
allows a direct coupling of the $\rho_G$ to the conserved QCD current,
of the form $\alpha\,\rho^a_\mu J^a_\mu$. Through the vector meson
dominance calculation, one can absorb this interaction into the
kinetic mixing term $\rho_G^{a\, \mu\nu} G^a_{\mu\nu}$ in
Eq.~(\ref{eq:leff2}).

We now comment on the mass ratio $m_{\pi_G}/m_{\rho_G}$.  The axial
$SU(3)$ subgroup of the global chiral flavor symmetry is explicitly
broken when the vector subgroup is identified with (gauged) QCD.  This
ensures that even in the absence of bare masses for $\Psi$, QCD
interactions will generate a mass $m_{\pi_G}$ for the $\pi_G$ octet.
The size of the generated $m_{\pi_G} $ relative to the cutoff, and in
particular relative to $m_{\rho_G}$, depends on the unknown strong
dynamics of $G$.  Previous studies have used QCD as a model to
calculate the mass ratio $m_{\pi_G}/m_{\rho_G} $, finding
\cite{Kilic:2008pm}
\beq
\frac{m_{\pi_G}^ 2}{m_{\rho_G}^ 2} = 3 \left(\frac{\alpha_s}{\alpha}\right)  
             \frac{\left. \delta m_{\pi} ^ 2\right|_{EM}}{m_{\rho} ^ 2}\simeq 0.3,
\eeq
based on the observed electromagnetic contribution to the pion mass
splitting $ \left.\delta m_{\pi} ^ 2\right|_{EM} \simeq \frac{3
\alpha}{4\pi} \, 2 \ln 2\, m_{\rho} ^ 2$. Again, this numerical result
depends on detailed properties of the QCD spectral functions whose
genericity is unclear.  A general estimate, 
not using the simple $N_c$ counting in QCD, suggests
the pNGB mass to scale like
\beq
m^2_{\pi_G} \sim \frac{g_s ^ 2}{(4\pi) ^ 2} \Lambda^2_G,
\eeq
where $\Lambda_G$ is the cutoff.  For $\rho_G$ with mass
of order the cutoff, we can then estimate
\beq
\frac{m_{\pi_G} }{m_{\rho_G}} \sim 0.1.
\eeq
The above estimation is based on naive dimensional analysis and some
order unity numbers can easily modify this relation, which depends on
the underlying strong dynamics.  Additional explicit sources of chiral
symmetry breaking would yield additional contributions to the $G$-pion
mass.  We focus our attention on the regime where $0.1\ltap
m_{\pi_G}/m_{\rho_G}\ltap 0.3$, where the $\pi_G$'s from $\rho_G$
decay are sufficiently boosted that searches will proceed more
profitably with jet substructure techniques.



\end{document}